\documentclass[journal]{IEEEtran}
\usepackage{multirow}
\usepackage{comment}
\usepackage{cite}
\usepackage{url}
\usepackage{amsmath,amsthm,amssymb,amsfonts}
\usepackage{algorithmic}
\usepackage{mathtools}
\usepackage{graphicx}
\usepackage{textcomp}
\usepackage{xcolor}
\usepackage{inputenc}
\usepackage{float}
\usepackage{physics}
\DeclareMathOperator{\E}{\mathbb{E}}
\usepackage{enumitem}
\newcommand{\bd}[1]{\mathbf{{#1}}}

\theoremstyle{definition} 
\newtheorem{theorem}{Theorem}

\newtheorem{corollary}{Corollary}
\newtheorem{lemma}{Lemma}
\newcommand{\paren}[1]{\left({#1}\right)}

\newcommand{\bracket}[1]{{\left [{#1}\right ]}}
\newcommand{\ith}[1]    {{#1}^{\underline{ \text{th}}}}
\newcommand{\ist}[1]    {{#1}^{\underline{ \text{st}}}}
\newcommand{\ind}[1]    {{#1}^{\underline{ \text{nd}}}}
\newcommand{\ird}[1]    {{#1}^{\underline{ \text{rd}}}}
\renewenvironment{cases}{\left\{\begin{array}[c]{ll}}{\end{array}\right.}
\begin{document}
\title{
Robust 2-D DOA Estimation in a Polarization Sensitive Single User Environment}
\author{Md~Imrul~Hasan,~\IEEEmembership{Member,~IEEE,}
        Mohammad~Saquib,~\IEEEmembership{Senior Member,~IEEE}
   
\thanks{ Md Imrul Hasan, and Mohammad Saquib are with the Department of Electrical and Computer Engineering, The University of Texas at Dallas, Texas, USA (e-mail: mxh170023@utdallas.edu, and saquib@utdallas.edu).}
}
\markboth{Journal of \LaTeX\ Class Files,~Vol.~X, No.~X, X~2022}%
{Md Imrul Hasan \MakeLowercase{\textit{et al.}}: Robust 2-D DOA Estimation in a Polarization Sensitive Single User Environment}
\maketitle
\begin{abstract}

Apart from the conventional parameters (such as signal-to-noise ratio, array geometry and size, sample size), several other factors (e.g. alignment of the antenna elements, polarization parameters) influence the performance of direction of arrival (DOA) estimating algorithms. When all the antenna elements are identically aligned, the polarization parameters do not affect the steering vectors, which is the underlying assumption of all the conventional DOA algorithms. Unfortunately, in this case, for a given set of DOA angles there exists a range of polarization parameters which could result in a very low signal-to-noise ratio (SNR) across all the antenna elements in the array. To avoid this type of catastrophic event, different antenna element needs to be aligned differently. However, this fact will make almost all commonly used DOA estimation algorithms non-operable, since the steering vectors are contaminated by the polarization parameters. To the best of our knowledge, no work in the literature addresses this issue even for a single user environment. In this paper, that line of inquiry is pursued. We consider a circular array with the minimum number of antenna elements and propose an antenna alignment scheme to ensure that at any given point no more than one element will suffer from significantly low SNR due to the contribution of polarization. A low complexity algorithm that estimates the DOA angles in a closed-form manner is developed. We treat MUSIC as the baseline algorithm and demonstrate how it can reliably operate in all possible DOA and polarization environments. Finally, a thorough performance and complexity analysis are illustrated for the above two algorithms.

\end{abstract}
\begin{IEEEkeywords}
Single source, DOA estimation, polarization, closed-form estimation, MUSIC, antenna alignment, antenna mapping.
\end{IEEEkeywords}
\IEEEpeerreviewmaketitle
\section{Introduction}
\IEEEPARstart{D}{irection} of arrival (DOA) is a key issue in various important applications, such as sonar, radar, medical sector, astronomy, defense operations, navigation, geophysics, acoustic tracking, and so on \cite{a01, a04, a05, a06, mp2}. Nowadays, owing to the exceptional development of modern technology and smart devices, DOA estimation techniques have been widely used in wireless communication and the internet of things (IoT) \cite{a07, a08, a09, a010, a011}. The localization of a single narrow-band source by a passive sensor array has also attracted tremendous interest in the literature due to its numerous applications \cite{a2,a3,a4,a5,ss1,ss2,ppp,low}. Over several decades, extensive studies have been performed, and numerous algorithms have been developed to estimate the DOAs, i.e., multiple signal classification (MUSIC) \cite{b1}, maximum-likelihood (ML) \cite{b2}, Capon \cite{b01}, estimation of signal parameters via rotational invariance techniques (ESPRIT) \cite{b4}, Min-Norm \cite{b5}, etc. Among all these methods, the MUSIC algorithm is especially noteworthy due to its easy implementation with different array structures. 

Antenna array geometry plays an important role while implementing the DOA estimation algorithms. To estimate the DOAs, linear array geometry is extensively studied in the literature \cite{l1,l2,l3}. However, the important drawback of a linear antenna array is the 1-D angle scanning. The circular antenna array (UCA) is offered to overcome the problem owing to its advantage of providing 360-degree azimuthal coverage as well as the elevation information of the DOAs \cite{lc1,lc2,hasan,lc3,lc4}.  
The polarization sensitivity of the array to incident signals needs to be considered in the DOA estimation. The received signal power is greatly affected by polarization while forming a transmitter-receiver pair. When the polarization of the receiver antenna matches with the transmitter antenna's polarization, the receiver collects the signal with the maximum possible power. On the other hand, a polarization mismatch between the transmitter-receiver pair can result in severe degradation of the received signal power. The polarization of the transmitter can be previously known; however, the polarization state of the transmitted signal can change when the electromagnetic wave scatters from a target. Therefore, keeping all the antenna elements in the same direction to match the polarization of the transmitter can often cause to receive the signal with a very low signal-to-noise ratio (SNR). Contrarily, using different directions for different antennas will affect the elements of the steering vector differently. This could lead the regular DOA estimation algorithms to exhibit unreliable performance. 

Two kinds of strategies have been used in the existing literature to eliminate the effects of polarization for DOA estimation algorithms. The works in the first category \cite{c1, c2, c3}  handle the problem utilizing complex hardware. In \cite{c1}, a uniform linear array with crossed dipoles was used for ESPRIT \cite{b4} to jointly estimate the DOAs and the polarization parameters. The MUSIC algorithm of joint polarization-DOA estimation based on the polarization-sensitive circular array with a crossed dipole is discussed in \cite{c2}. A uniform linear crossed tripole array is introduced for a dimension-reduction-based MUSIC algorithm in \cite{c3}. The works in the second category \cite{c7} managed the polarization issue by facing all the antenna elements towards the same direction which can cause them to suffer from low SNR due to the contribution of the polarization. A handful of works adopted both the categories at the same time, e.g. \cite{c4, c5} have performed the estimation using a circular vector sensor array comprising of co-centered orthogonal loop and dipole (COLD) pairs placing all of them in the same direction.

In this paper, we present how to localize a single narrowband source using a UCA of simple short dipole antenna elements just employing signal processing techniques.  Here, we want to utilize the minimum number of antenna elements, and at the same time, ensure that no more than one antenna element suffers from significantly low received power or SNR owing to the contribution of polarization. This fact leads us to mathematically develop an antenna alignment scheme, which also helps us to cancel the effects of polarization from the non-signal subspace while implementing the popular MUSIC algorithm. The entire algorithm is referred to as C-MUSIC. While cleansing the non-signal subspace for MUSIC, we develop a reduced complexity algorithm, namely CF, that estimates the DOA angles in a closed-form manner. The other major contributions of this paper can be outlined as follows:
\begin{enumerate}
\item As claimed, under the proposed antenna element alignment scheme, no more than one element will suffer from very low received power due to the polarization contribution. A decision threshold $K$ is introduced to decide whether the antenna element with the smallest received power should be considered in the process of DOA estimation or not.
\begin{enumerate}
\item  Two methods are developed to design the threshold $K$. The $\ist{1}$ method is based on the well-known Neyman-Pearson lemma and the $\ind{2}$ one is based on the central limit theorem (CLT).  
\item An extensive probabilistic analysis of $K$ is performed in order to provide significant insights into the impact of $K$ on the system performance. 
\end{enumerate}
\item Two non-signal subspace cleansing methods are demonstrated for C-MUSIC.
\begin{enumerate}
\item One of the methods can be employed only when the smallest received power among all the antenna elements is below the threshold $K$. The other one applies to all possible received power scenarios.
\end{enumerate}
\item A through performance and complexity study is performed between C-MUSIC and CF algorithms.  
\end{enumerate}
The rest of the paper is organized as follows:
Section~\ref{Model} presents the System Model. In Section III, the problem statement is formulated and the DOA estimation algorithms (C-MUSIC and CF) are developed in Section IV assuming no background noise. Implementations of those algorithms in a noisy arbitrarily polarized environment are discussed in Section V with analysis. In Section VI, the numerical results are presented, and finally, the concluding remarks of this paper are given in  Section VII. 

\textit{Notations:} We use lowercase and uppercase bold letters to denote vectors and matrices, respectively. Lowercase letters in italics are used to represent scalars. The notation * refers to the complex conjugate of a scalar, $[\cdot]^T$ refers to transpose, and $[\cdot]^\dag$ denotes the Hermitian of a matrix.

\section{System Model}\label{Model}
\begin{figure}[t!]
\centering
    \includegraphics[scale=0.85]{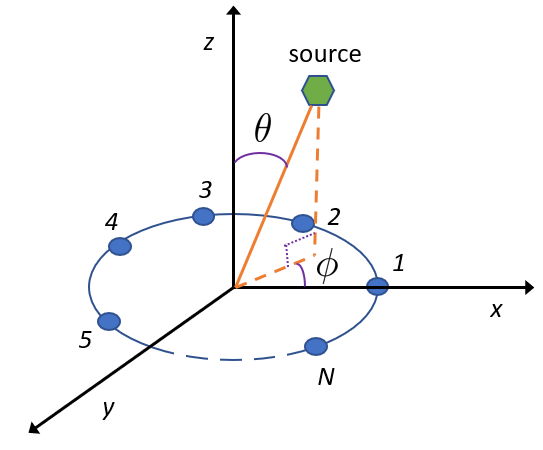}
    \caption{Uniform circular array geometry.}
    \label{fig:Capture1}
\end{figure}

Let's consider a UCA of radius $r$ composed of arbitrarily aligned $N$  identical short dipole antennas. The elements are positioned on the circumference of the circle on the $xy-$plane where the first element is located on the $x-$axis; see Fig. \ref{fig:Capture1}. These surface antenna elements are low profile, easily mountable, and very robust on a rigid surface \cite{w}.
The incident signal at the array is a complex unknown narrow-band signal $s(m)$ coming from a far field source. The azimuth angle $\phi$ of this signal is measured counterclockwise from the $x$ axis, and the elevation angle $\theta$ is measured downward from the $z-$axis; where $0^\circ\leq \phi< 360^\circ$ and  $0^\circ\leq\theta\leq 90^\circ$. The signal is arbitrarily polarized, where $0^\circ\leq \eta <360^\circ$, and  $0^\circ\leq \gamma < 90^\circ$ represent the polarization phase difference, and the auxiliary polarization angle, respectively. These terms satisfy $\tan\gamma={A_{y}}/{A_{x}}$, and $\eta= \phi_{y}-\phi_{x}$, where, $A_{y}$ and $\phi_{y}$ are the electric field amplitude and the electric field phase along the $y-$axis, respectively, and $A_{x}$, and $\phi_{x}$ are defined accordingly for $x-$axis \cite{c6}.

\begin{figure}[t!]
 \centering
 \includegraphics[scale=0.80]{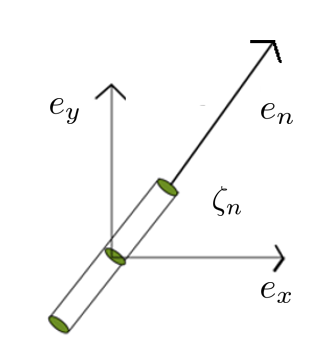}
    \caption{Electric field in short dipole antenna.}
    \label{fig:Capture2}
\end{figure}

 The output voltage from each short dipole is proportional to the electric field component along the dipole axis and the length of the dipole  \cite{r}. Hence, the outputs of these identical elements parallel to the $x$, and
$y$ axes will be proportional to the $x$, and $y$ components of the electric field,
respectively.
According to  \cite{r,c5}, the electric fields along $x$, and $y$ axes are given by
 \begin{equation}
\begin{bmatrix}
e_\mathrm{x}   \\
 e_\mathrm{y}   
\end{bmatrix}= \begin{bmatrix}
e^{j\eta} \sin \gamma \cos \theta \cos \phi -\sin \phi \cos \gamma \\
e^{j\eta} \sin \gamma \cos \theta \sin \phi +\cos \phi \cos \gamma 
\end{bmatrix}\,.
\label{a00}
\end{equation}
At the $\ith{n}$ dipole antenna element, the output voltage is proportional to the electric field
\begin{align}
  \label{a1}
    e_{n} &=e^{j\eta} \sin \gamma \cos \theta \cos (\phi-\zeta_n) -\sin (\phi-\zeta_n) \cos \gamma \nonumber \\ 
    &=e_\mathrm{x} \cos \zeta_n+ e_\mathrm{y} \sin \zeta_n\,,
\end{align}
where $\zeta_n$ is the alignment angle of that element w.r.t. the $x-$axis; see Fig. \ref{fig:Capture2}. Notice in (\ref{a1}) that the electric field components $e_\mathrm{x}$, and $e_\mathrm{y}$ of (\ref{a00}) are special cases of $e_n$ for $\zeta_n=0^{\circ}$, and $\zeta_n=90^{\circ}$, respectively. Equation (\ref{a1}) also implies that for $\theta= 90^{\circ}$, and $\gamma= 90^{\circ}$, irrespective of any antenna alignment angle $\zeta_n$, the received signal power (RSP) = 0 at all the  antenna elements due to the electric field $e_n=0$. If that is the case, we decide the source is along the $xy$-plane, since $\theta= 90^{\circ}$, and the value of $\phi$ can not be determined. This situation can be overcome by having one or more short dipole antenna elements along the $z-$axis at the cost of losing low profile and very robust properties of the surface antenna elements \cite{w}, which is beyond the scope of this paper. Throughout the rest of this paper, we consider $\theta\neq 90^{\circ}$ or $\gamma \neq 90^{\circ}$ which is equivalent to $e_x\neq 0$ or $e_y\neq 0$. 

The steering term for the $n^{\underline{\mathrm{th}}}$ element shown in Fig.~\ref{fig:Capture1} relates to the source's DOA angles $(\theta,\phi)$ as
\begin{equation}
    {a_{\mathrm{s}_n}}=e^{j\frac{2\pi}{\lambda}r\sin\theta \cos(\phi-\beta_n)} ,
    \label{a_s}
\end{equation}
where $\beta_n ={2\pi(n-1)}/{N}$; $n=1,2,3,...,N$, and $\lambda$ denotes the wavelength of the target signal. 
The array manifold term for the  $n^{\underline{\mathrm{th}}}$ antenna element can be written as the  product of the associated electric field response and the steering terms \cite{c4,c5,c6} as
\begin{equation}
 a_n=  e_{n} a_{\mathrm{s}_{n}}.
 \label{a_nn}
\end{equation}

At the $m^{\underline{\mathrm{th}}}$ snapshot, the output of element $n$ is
  \begin{equation}
 {{ { {x}}}_n}(m)= {a_n}s(m)+w_n(m)\,,
 \label{a_n0}
 \end{equation}
 where $w_n(m)$ is zero-mean white complex Gaussian noise with average power $\sigma^2$, and spatially and temporally independent of $s(m)$. 
 
 Now collecting outputs from all $N$ antenna elements, we form:
   \begin{equation}
    \mathbf{{x}}(m)=\mathbf{a}{s(m)}+\mathbf{w}(m),
    \label{RSE}
 \end{equation}
where
 \begin{align}
      \mathbf {a}=\begin{bmatrix}
 a_{{1}}  \\
 a_{{2}}   \\
\vdots\\
 a_{{N}}  
\end{bmatrix};
\mathbf{{x}}(m)=\begin{bmatrix}
{x}_1(m)\\
{x}_2(m)\\
\vdots\\
{x}_N(m)
\end{bmatrix};
  \mathbf{w}(m)=\begin{bmatrix}
{w}_1(m)\\
{w}_2(m)\\
\vdots\\
{w}_N(m)
\end{bmatrix} .
\label{a_8}
 \end{align}
Equation (\ref{RSE}) will be processed to extract the desired DOA information using C-MUSIC and CF algorithms. Both the algorithms will be developed soon assuming no background noise (i.e., $\sigma= 0$). Subsequently, we will show how those algorithms will be implemented in real-life scenarios in the presence of background noise. Before we do so, the importance of mitigating the polarization contribution will be signified in the context of the conventional MUSIC algorithm.

\section{Problem Formulation: Limitations of MUSIC}
Let's begin with a short description of the subspace based 2-D MUSIC algorithm. It exploits the eigen structure of the auto-correlation matrix of the received signal (\ref{RSE}), which is
 \begin{align}
 \mathbf{R}& =\E
 \{\mathbf{{x}}(m)\mathbf{{x}}(m)^\dag\}\nonumber\\
 &=\mathbf{a} s(m)s(m)^\dag\mathbf{a}^\dag+\E \{\mathbf{w}(m)\mathbf{w}(m)^\dag\}\nonumber\\
 &=\sigma^2_{\mathrm{s},m}\mathbf{a}\mathbf{a}^\dag+\sigma^2\mathbf{I}\,,
 \end{align}
 where $\sigma^2_{\mathrm{s},m}=|s(m)|^2$ ,  $\mathbb{E}\{\mathbf{w}(m)\mathbf{w}(m)^\dag\} = \sigma^2 \mathbf{I}$, and  $\mathbf{I}$ is an $N\times N$ identity matrix. In practice, the above auto-correlation matrix will be replaced by the sample auto-correlation matrix averaged over $M$ time samples (or snapshots). Since a single source is considered, the largest eigenvalue of this autocorrelation matrix is corresponding to the signal subspace. The others are corresponding to the non-signal (or noise) subspace which is defined as
\begin{equation}
  \mathbf{E}= [\mathbf{v}_2,\mathbf{v}_3,\cdot \cdot \cdot, \mathbf{v}_N ]\,,
\label{P_M1}
\end{equation}
where  $\{\mathbf{v}_l\}_{l=1}^{N}$ denotes the eigenvector corresponding to the real eigenvalue $\{\lambda_l\}_{l=1}^{N}$ of  $\mathbf{R}$. These eigenvalues are sorted in the descending order. Using the noise subspace $\mathbf{E}$ (\ref{P_M1}),  the MUSIC spectrum is defined as \cite{a9}
\begin{equation}
    {S_\mathrm{MUSIC}}=\frac{1}{\mathbf{a}^\dag \mathbf{E} {\mathbf{E}}^\dag \mathbf{a}}\,.
    \label{P_M}
\end{equation}
Next, we discuss the implementation requirement of the MUSIC algorithm in an arbitrarily polarized environment. MUSIC needs at least two eigenvectors in the noise subspace in order to estimate two DOA angles (azimuthal and elevation)  unambiguously. Therefore, the minimum required number of antenna elements is three.  Due to the polarization and DOA angles, the compound steering element $|a_n|$ may be approximately $0$ in (\ref{a_nn}) at one (or more) antenna depending on the alignment resulting in very low signal power. Following is an example of this catastrophic situation.

Let's assume all three antennas are aligned along the $x-$ axis, and the incoming wave arrives with the azimuthal angle $\phi=90^{\circ}$, and the auxiliary polarization angle  $\gamma=90^{\circ}$. This set of parameters yields received electric field $e_n=e_{\rm{x}}=0$, hence $a_n=0$, for all $n$ which leads the received signal in (\ref{a_8}) to be zero in the noiseless scenario. In this situation, no regular DOA estimating algorithm will be able to operate.
To resolve this issue, one must use \underline{at least four element UCA} with \underline{different alignments for different elements} such that the above harmful event (i.e. $|a_n|\approx 0$) can't affect more than one antenna element. 

\begin{figure}[t!]
   \includegraphics[scale=0.60]{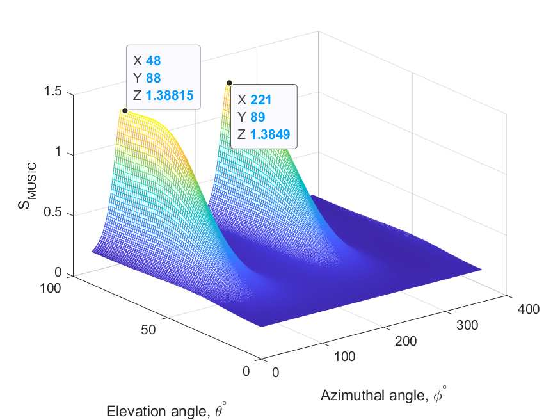}
    \caption{MUSIC spectrum in a polarization sensitive environment.}
    \label{fig:Capture5}
\end{figure}
Unfortunately, a four element UCA with different alignments is not even enough for the conventional MUSIC to operate. It is due to the fact that the conventional MUSIC assumes $e_n=e_{n'}$ for $n\neq n'$ in (\ref{a1}), that results  the  non-signal subspace free from the contribution of the polarization parameters. It is not true when the antenna elements are aligned differently. At the $n^{\underline{\mathrm{th}}}$ element, the compound term in (\ref{a_nn}) can be written as
\begin{equation}
     a_n= e_{n} a_{s_{n}}=|{e_{n}}|e^{j\delta_n} |a_{s_{n}}|e ^{j\psi_n}\,,
     \label{a_n}
\end{equation}
where $|a_{s_{n}}|=1$. Terms $\delta_n$ and $\psi_n$ are the phases of voltage $e_{n}$, and  steering element $a_{s_{n}}$, respectively. The element $(p,q)$ of the auto-correlation matrix $\mathbf{R}$ is $\sigma^2_{\mathrm{s},m} |e_p||e_q|e^{j(\delta_{p,q}+\psi_{p,q})}$, where $\delta_{p,q}=\delta_{p}-\delta_{q}$, and $\psi_{p,q}=\psi_{p}-\psi_{q}$. Since all the antenna elements have different alignments, $e_{p} \neq e_{q}$ for $p\neq q$ and therefore, different elements of the auto-correlation matrix $\mathbf{R}$ are affected by the polarization parameters differently. This forces the associated non-signal subspace to deviate from the desired one, which is free from the polarization parameters. This could cause failure of the MUSIC algorithm; see Fig. \ref{fig:Capture5}, where MUSIC spectrum is plotted using the DOA angle pair, $\theta=30^{\circ}$, $\phi=60^{\circ}$, and the polarization parameter pair, $\gamma=45^{\circ}$, $\eta=45^{\circ}$, and alignment angle of antenna element $n$, $\zeta_n= (n-1)\times 30^{\circ}$, where $n=1,2,3,4$. More discussion and analysis on this issue will be found in the following section. Note that MUSIC performs perfectly for the above set of DOA and polarization angles, if all the antenna elements are aligned in the same direction (i.e., $\zeta_n = \zeta_{n'}$ for $n\neq n'$). However, this same alignment could result the catastrophic incident for a different set of DOA and polarization angles as demonstrated in the earlier example.

The above two examples jointly suggest that in a single user arbitrarily polarized environment to find the DOA angles, different alignment must be used for different elements, and the implementation of MUSIC requires cleansing of the polarization parameters from the non-signal subspace. Such schemes will be developed next considering all possible system conditions. Thus, in this work MUSIC is referred to as \textbf{C-MUSIC}, where the alphabet `C' stands for the additional signal processing task (cleansing operation of the polarization contribution)  performed prior to implementing the conventional MUSIC algorithm. In the following section, we will also demonstrate that after cleansing, how one can estimate the DOA angles in a closed-form manner using the CF algorithm.

\section{Estimation of the DOA Angles: C-MUSIC and CF}
Before start elaborating upon the C-MUSIC and CF algorithms, let's derive the antenna alignment guideline to ensure that the effects of the polarization should not be able to hurt more than one antenna element in terms of RSP as long as $e_x\neq 0$ or $e_y\neq 0$. In that regard, the following theorem would be useful.   
 
\begin{theorem}
If antenna elements $n$ and $n' \neq n$ are aligned such that $\zeta_{n'} \neq \zeta_n$ or $ \zeta_{n'} \neq \zeta_n \pm 180^\circ$, the RSP can't be simultaneously zero at  antenna $n$ and $n'$ when $e_x\neq 0$ or $e_y\neq 0$.
\label{Claim-1}
\end{theorem}
\begin{IEEEproof}
Let's assume RSP = 0 at antenna element $n$ and equivalently from (\ref{a1}), we get the received electric field at antenna $n$
\begin{equation}
    e_{n}=e_x \cos \zeta_n+ e_y \sin \zeta_n=0\,. 
    \label{a2}
\end{equation}

Without loss of generality, we can write the alignment angle of antenna $n'\neq n$ as 
\begin{equation}
\zeta_{n'} =\zeta_{n} + \zeta\,,
\label{a31}
\end{equation}
where $0\leq \zeta < 360^\circ$. Applying the above equation and (\ref{a2}) to (\ref{a1}), and after simplification, the received electric voltage at antenna $n'$ becomes
\begin{equation}
    e_{n'}=  \sin\zeta\paren{e_x \sin \zeta_n- e_y \cos \zeta_n}\,. 
    \label{a3}
\end{equation}

If $e_x\neq 0$ or $e_y\neq 0$, (\ref{a2}) implies
$$e_x \sin \zeta_n- e_y \cos \zeta_n \neq 0\,,$$
equivalently 
$$e_{n'}\neq0\,,$$
 unless $\zeta \neq 0$ (i.e., $\zeta_{n'} \neq \zeta_n$ ) or $\pm 180^\circ$ (i.e., $ \zeta_{n'} \neq \zeta_n \pm 180^\circ$) in (\ref{a31}). This proves the theorem.
 \end{IEEEproof}
 
\begin{figure}[t!]
    \includegraphics[width=\columnwidth]{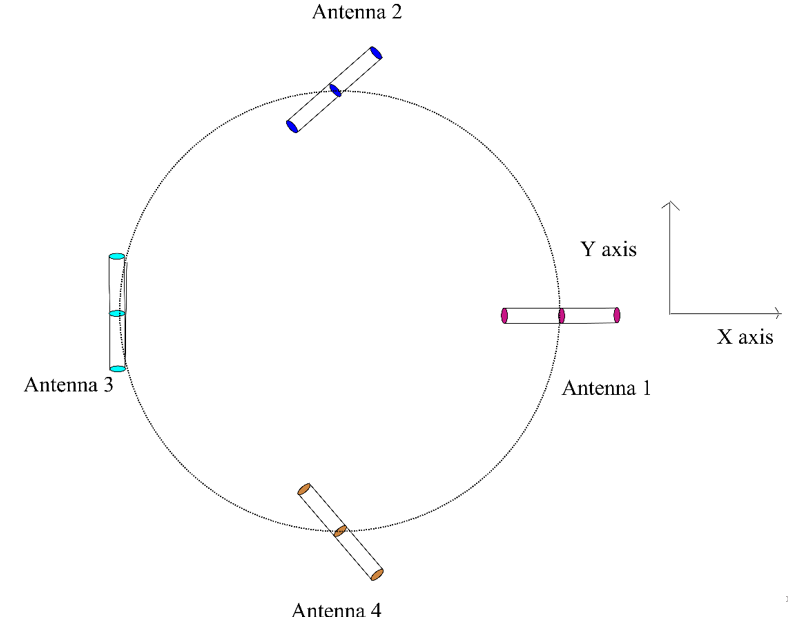}
    \caption{Proposed antenna alignment scheme for a UCA with $N=4$.}
    \label{fig:Capture6}
\end{figure}

Condition  $\zeta_{n'} \neq \zeta_n$ for $n\neq n'$ in Theorem~\ref{Claim-1} implies that all 4 antenna elements must be aligned differently. In addition, if we choose the antenna element alignment such that $0^{\circ}\leq \zeta_n < 180^\circ$ for all $n$, the other requirement $ \zeta_{n'} \neq \zeta_n \pm 180^\circ$ in Theorem~\ref{Claim-1} will be met. One such array that meets both criteria of Theorem~\ref{Claim-1} is depicted in Fig. \ref{fig:Capture6}. Here, it can be seen that antenna 1 and 3 are aligned along the $x$- and $y$- axis, respectively. Antenna 2 and 4 are aligned by $45^{\circ}$, and $135^{\circ}$, respectively, with respect to the $x$-axis. C-MUSIC and CF algorithms now will be developed using the above antenna element alignment scheme.  

\subsection{C-MUSIC}
Theorem~\ref{Claim-1} and the antenna alignment scheme given in Fig. \ref{fig:Capture6} jointly result in five different RSP scenarios with the presence of a target. We categorize those scenarios into two primary events. In Event 1, $a_n= 0$ only at antenna element $n$; since $n=1,2,3,4$, there are four sub-events. Term $\Omega_{1,n}$ represents the ${\ith{n}}$ sub-event of Event 1.  In Event 2 (denoted by $\Omega_2$),  $a_n\neq 0$ at all the antenna elements.  In practice, Event 2 occurs much more frequently than Event 1, since the former to happen requires a special combination of polarization parameters ($\eta$ and $\gamma$) for a given pair of DOA angles ($\phi$ and $\theta$). Nevertheless, the DOA estimation algorithms must operate efficiently in each of the above possible events. Structures of the signal and non-signal subspace will vary from one event to another. This fact requires a dedicated polarization suppression method for each of the five effective events. As a result, to apply the right method, the event that has occurred must be correctly identified. For the purpose of the demonstration, we assume the perfect detection of the events given the RSP scenarios at all the antennas. Later, we will present two threshold-based techniques to operate the algorithm in a real-life scenario. Let's begin with Event 1 and demonstrate how to implement C-MUSIC.

  \subsubsection{Event 1} Firstly, we consider the first sub-event of event 1, i.e., $\Omega_{1,1}$. At the $\ith{m}$ sample,  for $\beta_n ={2\pi(n-1)}/{N}$ and $\zeta_n=(n-1)\times 45^{\circ}$, the signal component of the received signal (\ref{RSE})  is  
\begin{equation}
   \bd{x}(m) = \mathbf{s}_{\mathrm{t}}s(m)= \begin{bmatrix}
0\\
\frac{1}{\sqrt2}  e_y e^{j\kappa_2}\\
e_y  e^{-j\kappa_1}\\
\frac{1}{\sqrt2} e_y e^{- j\kappa_2}
\end{bmatrix}s(m)\,,
\label{a_201}
\end{equation}
where $\kappa_1 = \kappa \cos \phi$, $\kappa_2 = \kappa \sin \phi$ and $\kappa = 2\pi r\sin\theta/\lambda$. Removing the first element (i.e., $0$) from the original received signal vector, we form a new received signal vector as   
 \begin{equation}
   \bd{\tilde{x}}(m) =  e_y\mathbf{\tilde{s}}_{\mathrm{t}}s(m),  \,
\label{a_202}
\end{equation}
where $ \mathbf{\tilde{s}}_{\mathrm{t}} = \begin{bmatrix}
\frac{1}{\sqrt2}  e^{j\kappa_2}\quad
 e^{-j\kappa_1}\quad
\frac{1}{\sqrt2} e^{- j\kappa_2}
\end{bmatrix}^\top\,.$
Note that in the newly formed received signal vector (\ref{a_202}), the polarization contributes two multiplicative terms, one is $e_y$, and the other one is $1/\sqrt 2$ (in the $\ist{1}$ and $\ird{3}$ elements of vector $\mathbf{\tilde{s}}_{\mathrm{t}}$). Vector $\mathbf{\tilde{s}}_{\mathrm{t}}$ carries useful information pertaining to the DOA. In event 1, C-MUSIC can be implemented in two ways, the first method, namely Method 1, executes by cleaning the non-signal subspace of the auto-correlation matrix of the received signal (\ref{a_202}) and, the second method, namely Method 2, operates estimating $\kappa_1$ and $\kappa_1$. Next we describe Method 1.

\paragraph{Method 1}
We can write the auto-correlation matrix of the received signal as
  \begin{equation}
  \bd{\tilde{R}} = |e_y|^2\sigma_{\mathrm{s},m}^2 \mathbf{\tilde{s}}_{\mathrm{t}} \mathbf{\tilde{s}}_{\mathrm{t}}^\dag  = |e_y|^2\sigma_{\mathrm{s},m}^2 \bd{\tilde{R}}_{\mathrm{s}},
  \label{R1} 
 \end{equation}
 where  $$\bd{\tilde{R}}_{\mathrm{s}} = \begin{bmatrix}
\frac{1}{2} & \frac{1}{\sqrt 2}e^{j(\kappa_2+\kappa_1)} & \frac{1}{2}e^{2j\kappa_2}  \\
\frac{1}{\sqrt 2} e^{-j(\kappa_1+\kappa_2)}  & 1 &\frac{1}{\sqrt 2} e^{-j(\kappa_1-\kappa_2)} \\
\frac{1}{2}e^{-2j\kappa_2} &\frac{1}{\sqrt 2}e^{j  (\kappa_1-\kappa_2)} & \frac{1}{2} \end{bmatrix} \,.$$

To develop C-MUSIC, we derive a subspace from $  \bd{\tilde{R}}$ identical to the non-signal subspace of
$$\bd{\tilde{R}}_{\mathrm{s,c}} =  \mathbf{\tilde{s}}_{\mathrm{t,c}}  \mathbf{\tilde{s}}_{\mathrm{t,c}}^\dag,$$
where $\mathbf{\tilde{s}}_{\mathrm{t,c}} =  \begin{bmatrix}
e^{j\kappa_2}\quad  
 e^{-j\kappa_1}\quad
 e^{- j\kappa_2}
\end{bmatrix}^\top$ is the clean version of the compounded steering vector $\mathbf{\tilde{s}}_{\mathrm{t}}$ in (\ref{a_202}). The above auto-correlation matrix $\bd{\tilde{R}}_{\mathrm{s,c}} $ can be shown as
\begin{equation}
    \bd{\tilde{R}}_{\mathrm{s,c}}= \begin{bmatrix}
1 & e^{j(\kappa_2+\kappa_1)} & e^{2j\kappa_2}  \\
 e^{-j(\kappa_1+\kappa_2)}  & 1 &  e^{-j(\kappa_1-\kappa_2)} \\
e^{-2j\kappa_2} & e^{j(\kappa_1-\kappa_2)} & 1 \end{bmatrix} \,.
\label{rsc}
\end{equation}
We now prove the following theorem  to derive the non-signal subspace of $\bd{\tilde{R}}_{\mathrm{s,c}}$ from  $\bd{\tilde{R}}$ of (\ref{R1}).

\begin{theorem}
The non-signal subspace of $\bd{\tilde{R}}_{\mathrm{s,c}}$ in (\ref{rsc}) is the null space of $\bd{\tilde{R}}\bd{F}_1$, where $  \mathbf{F}_1=\mathrm{diag}\bracket{1 \quad \frac{1}{\sqrt 2}\quad   1 }$. 
\label{claim-Method 1}
 \end{theorem}
 \begin{IEEEproof}
 Let $\bd{v} =\begin{bmatrix}v_1\quad v_2\quad v_3\end{bmatrix}^\top $ be a vector belonging  to the non-signal subspace of $\bd{\tilde{R}}_{s,c}$, then
\begin{equation}
  \begin{bmatrix}
1 & e^{j(\kappa_2+\kappa_1)} & e^{2j\kappa_2}  \\
 e^{j(\kappa_1+\kappa_2)}  & 1 &  e^{-j(\kappa_1-\kappa_2)} \\
e^{-2j\kappa_2} & e^{j(\kappa_1-\kappa_2)} & 1 \end{bmatrix} 
 \begin{bmatrix}v_1\\v_2\\v_3\end{bmatrix} =\begin{bmatrix}0\\0 \\0\end{bmatrix}\,. \label{op}
 \end{equation}
Since $\bd{\tilde{R}}\bd{F}_1\bd{v}$ equals to
 \begin{equation*}
|e_y|^2\sigma_{\mathrm{s},m}^2  \begin{bmatrix}
\frac{1}{2} & \frac{1}{2}e^{j(\kappa_2+\kappa_1)} & \frac{1}{2}e^{2j\kappa_2}  \\
\frac{1}{\sqrt 2} e^{-j(\kappa_1+\kappa_2)}  & \frac{1}{\sqrt 2} &\frac{1}{\sqrt 2} e^{-j(\kappa_1-\kappa_2)} \\
\frac{1}{2}e^{-2j\kappa_2} &\frac{1}{2}e^{j  (\kappa_1-\kappa_2)} & \frac{1}{2} \end{bmatrix} 
\begin{bmatrix}v_1\\v_2 \\v_3\end{bmatrix}\,,
   \end{equation*}
   applying (\ref{op}) to the above expression, we get $$\bd{\tilde{R}}\bd{F}_1\bd{v}=\begin{bmatrix}0\quad 0 \quad 0\end{bmatrix}^\top,$$
 and we prove the theorem. 
 \end{IEEEproof}
 Next, we find the angles corresponding to the maximum spectrum value in (\ref{P_M}) utilizing the null space of $\bd{\tilde{R}}\bd{F}_1$ and use those as the DOA estimates. Theorem~\ref{claim-Method 1} guides us how to obtain the non-signal subspace for implementing C-MUSIC when $a_n =0$ for $n=1$. Similarly, when $a_n = 0$ for $n\neq 1$, the auto-correlation matrix $\bd{\tilde{R}}$ will be formed discarding the received signal samples from that antenna element. As a result, the structure of $\bd{\tilde{R}}$ changes with $n$ that change the structure of $\mathbf{F}_n$ as follows:
$$\mathbf{F}_2 = \mathrm{diag}\bracket{1\quad -1\quad -\frac{1}{\sqrt 2}}; \quad \mathbf{F}_3 = \mathrm{diag}\bracket{1\quad \sqrt 2\quad -\sqrt 2};$$ and  $$\mathbf{F}_4 =\mathbf{F}_1\,.   $$
The derivations of $\{\bd{F}_n\}_{n=2}^{4}$ are similar to $\mathbf{F}_1$ and omitted for conciseness. Let's focus again on antenna when RSP = 0 at $n=1$, and elaborate upon Method 2.
 \paragraph{Method 2} The following lemma, which can be straightforwardly obtained from (\ref{a_201}), forms the basis for this method. 
 \begin{lemma}
The phase parameters $\{\kappa_i\}_{i=1}^2$ of the steering elements can be derived from each sample of the received signal vector $\bd{\tilde{x}}(m)$ in (\ref{a_201}) as $$\kappa_2 = \frac{1}{2}\angle \{{x}_{2}(m){x}^*_{4}(m) \},$$ and$$ \kappa_1 =  \angle \{{x}_{2}(m){x}^*_{3}(m)\}- \kappa_2 \,.$$
\label{lemma2a}
\end{lemma}
In a noisy environment, applying Lemma \ref{lemma2a} to each sampled received signal, and then averaging those over $M$ samples, we get the estimates of $\kappa_1$ and $\kappa_2$ which are denoted by  $\hat{\kappa}_1$ and $\hat{\kappa}_2$, respectively. The auto-correlation matrix $\bd{\tilde{R}}_{\mathrm{s,c}}$ (\ref{rsc}) can be estimated using those estimates. After that MUSIC algorithm can be readily applied.\footnote{Note that an auto-correlation matrix of size $4\times 4$ can be formed by padding $e^{j\hat{\kappa}_1}$ at the top of the newly estimated steering vector $\mathbf{\tilde{s}}_{\mathrm{t,c}}$. However, this padding will add complexity to the MUSIC algorithm.} 

When $a_n = 0$ for $n\neq 1$, the estimation procedure of $\kappa_1$ and $\kappa_2$ changes as given in the following lemma:

\begin{lemma}
For event $\Omega_{1,2}$ (i.e., when $a_n = 0$ for $n=2$), the  phase parameters $\{\kappa_i\}_{i=1}^2$ satisfy

$$\kappa_1 = \frac{1}{2}\angle \{-{x}_{1}(m){x}^*_{3}(m) \},$$ and$$ \kappa_2 =  \angle \{-{x}_{1}(m){x}^*_{4}(m)\}- \kappa_1 \,.$$

For event $\Omega_{1,3}$ (i.e., when $a_n = 0$ at $n=3$), the  phase parameters $\{\kappa_i\}_{i=1}^2$ satisfiy
$$\kappa_2 = \frac{1}{2}\angle \{-{x}_{2}(m){x}^*_{4}(m) \},$$ and$$ \kappa_1 =  \angle \{{x}_{1}(m){x}^*_{2}(m)\}+ \kappa_2 \,.$$
For event $\Omega_{1,4}$ (i.e., when $a_n = 0$ at $n=4$), the  phase parameters $\{\kappa_i\}_{i=1}^2$ satisfy

$$\kappa_1 = \frac{1}{2}\angle \{{x}_{1}(m){x}^*_{3}(m) \},$$ and$$ \kappa_2 =  \angle \{{x}_{2}(m){x}^*_{1}(m)\}+ \kappa_1 \,.$$
\end{lemma}
\subsubsection{Event 2}
 Now, we consider the event where $a_n \neq 0$ for all $n$, i.e., $\Omega_{2}$.
The contribution of the target in the received signal  (\ref{RSE})  at the $\ith{m}$ sample is  
\begin{equation}
   \bd{x}(m) = \mathbf{s}_{\mathrm{t}}s(m),
\label{a_20}
\end{equation}
where the joint contribution of the steering vector and the polarization parameters is embedded in
\begin{equation}
   \mathbf{s}_{\mathrm{t}}=\begin{bmatrix}
e_xe^{j\kappa_1}\\
\frac{1}{\sqrt2}  (e_x+e_y) e^{j\kappa_2}\\
e_y  e^{-j\kappa_1}\\
\frac{1}{\sqrt2} (-e_x+e_y)  e^{- j\kappa_2}
\end{bmatrix}\,.
\label{a_19}
\end{equation}
We want vector $\mathbf{s}_{\mathrm{t}}$  (\ref{a_19}) to be free from terms $e_x$ and $e_y$ as follows:
\begin{equation}
   \mathbf{s}_{\mathrm{t,c}} =\begin{bmatrix}
e^{j\kappa_1}\\
e^{j\kappa_2}\\
 e^{-j\kappa_1}\\
 e^{- j\kappa_2}\
\end{bmatrix}\,.
\label{a_191}
\end{equation}
 Using (\ref{a_19}), it can be shown that unlike Event 1, the auto-correlation matrix of $\bd{x}(m)$  is a joint function of $e_x$ and $e_y$ and
 the implementation of Method 1 in this case requires the knowledge of their ratio. This fact makes Method 1 very difficult to implement for Event 2, whereas, the principle of implementation of Method 2 remains unchanged. Similar to Event 1, Method 2 operates estimating $\{\kappa_i\}_{i=1}^{2}$. However, the estimation technique is different now, since the received signal vector $ \bd{x}(m)$ in (\ref{a_20}) is different than that in Event 1.  After estimating $\{\kappa_i\}_{i=1}^{2}$, a clean steering vector $ \mathbf{s}_{\mathrm{t,c}}$ will be formed (\ref{a_191}). The non-signal subspace of its auto-correlation matrix will be used for C-MUSIC. 

To estimate $\{\kappa_i\}_{i=1}^{2}$, we need to prove the following two lemmas.
\begin{lemma}
If 
\begin{equation*}
c_1(m)=x_1(m)x^*_2(m)+x^*_3(m)x_4(m)\,,
\end{equation*}
where $x_i(m)$ is the $\ith{i}$ element of $\bd{x}(m)$ (\ref{a_20}), then
\begin{equation}
\angle c_1(m)  = \kappa_1 - \kappa_2.
  \label{a10}
\end{equation}
\label{lemma1}
\end{lemma}
\begin{IEEEproof}
Equation (\ref{a_20}) yields 
 \begin{equation*}
 x_1(m)x^*_2(m)=\frac{1}{\sqrt 2} \paren{|e_x|^2+e_xe^*_y}e^{j(\kappa_1-\kappa_2)}\sigma_{s,m}^2,  
\end{equation*}
and 
\begin{equation*}
x^*_3(m)x_4(m)=\frac{1}{\sqrt 2}\paren{-e_xe^*_y+|e_y|^2}e^{j(\kappa_1-\kappa_2)} \sigma_{s,m}^2.
\end{equation*}
Now we add the above two equations to prove the lemma.
\end{IEEEproof}

\begin{lemma}
If 
\begin{equation*}
c_2(m)=-x_1(m)x^*_4(m)+x^*_3(m)x_2(m)\,,
\end{equation*}
then
\begin{equation}
 \angle c_2(m)  = \kappa_1+ \kappa_2\,.
 \label{a_9}
\end{equation}
\label{lemma2}
\end{lemma}
The proof of the above lemma is similar to that of Lemma~\ref{lemma1} and due to the conciseness, it is omitted. The above two lemmas straightforwardly yield the estimates of $\kappa_1$ and $\kappa_2$ for each sample $m$. After processing $M$ samples, the auto-correlation matrix for the steering vector $\mathbf{s}_{\mathrm{t,c}}$ can be formed to implement MUSIC.\footnote{To estimate both the DOA angles, we need at least a steering vector of length 3. Therefore an auto-correlation matrix of size $3\times 3$ can be formed by ignoring one of the elements of the newly estimated steering vector $\mathbf{s}_{\mathrm{t,c}}$ and it will save some computational complexity.} Now, we will show how the CF algorithm will be implemented using the estimates of $\kappa_1$ and $\kappa_2$.
\subsection{CF}
Recall that  $$\kappa_1 = \kappa \cos \phi;\quad \kappa_2 = \kappa \sin \phi;$$ and \begin{equation}\kappa = \sqrt {\kappa_1^2+\kappa_2^2}=2\pi r\sin\theta/\lambda \,.\label{asd}\end{equation} 
 
In (\ref{asd}), parameter $\kappa=0$ or $\kappa\neq 0$. 
\begin{enumerate}
    \item When $\kappa=0$, the elevation angle $\theta = 0^\circ$ indicating that the target is along the $z-$axis and the azimuthal angle has no significance. 
    \item When $\kappa\neq 0$, parameter $\kappa_1$ and $\kappa_2$ can't be simultaneously zero due to the elevation angle $0< \theta \leq 90^\circ$ in (\ref{asd}). The above 3 expressions will be used to estimate DOA angles in a closed-form manner considering following 3 combinations of the estimated $\kappa_1$ and $\kappa_2$. 
\begin{enumerate}
 \item\label{CFA} If ${\kappa}_1\neq 0$ and ${\kappa}_2\neq 0$, we estimate the azimuthal angle
 \begin{equation*}
     {\phi}=\arctan\{ {\kappa}_2/ {\kappa}_1  \}\,,
     \label{t1}
 \end{equation*}
and since $\kappa>0$, the sign of ${\kappa}_1$ or  ${\kappa}_2$ will help to estimate $\phi$ without ambiguity. We now use (\ref{asd}) to estimate of the elevation angle as $$ {\theta} = \arcsin\paren{{\kappa}\lambda/2\pi r}\,. $$

\item If ${\kappa}_1 = 0$ and ${\kappa}_2 \neq 0$, the estimate of the azimuthal angle is either $90^\circ$ (for ${\kappa}_2 >0$) or $270^\circ$ (for ${\kappa}_2 < 0$). The estimate of term $\kappa$ is $|{\kappa_2}|$ which yields the estimate of the elevation angle as
 $$ {\theta} = \arcsin\paren{| {\kappa}_2|\lambda/2\pi r}\,.$$
 
 \item If ${\kappa}_1 \neq 0$ and ${\kappa}_2 = 0$, the estimate of the azimuthal angle is either $0^\circ$ (for ${\kappa}_1 >0$) or $180^\circ$ (for ${\kappa}_1 < 0$). The estimate of term $\kappa$ is $|{\kappa_1}|$ which yields the estimate of the elevation angle as
 $${\theta} = \arcsin\paren{|{\kappa}_1|\lambda/2\pi r}\,.$$
 \end{enumerate}
\end{enumerate}
Next, we demonstrate how the above two algorithms (C-MUSIC and CF) will be implemented in the presence of background noise. Here, the estimates of $\kappa_1$ and $\kappa_2$ may not be zero and the CF algorithm will execute following the above conditions given in~\ref{CFA}.

\begin{figure}[t!]
    \includegraphics[scale=0.59]{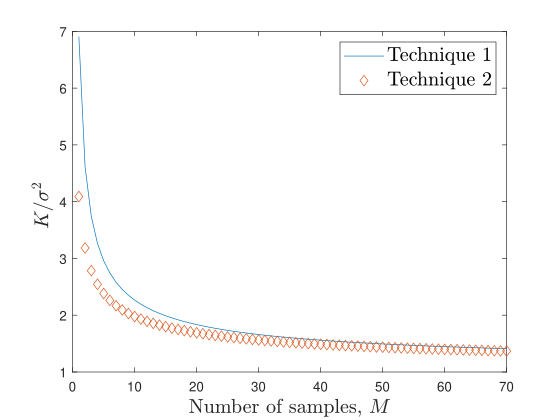}
    \caption{Normalized threshold $K$ vs number of samples $M$.}
    \label{fig:K}
\end{figure}
\section{Implementation of the Algorithms and Analysis}
In the above section, the working principles of the C-MUSIC and CF algorithms were described using the following two assumptions:
\begin{enumerate}
    \item {\bf Assumption 1:} The system is \underline{noise free} (i.e., $\sigma^2=0$).
     \item {\bf Assumption 2:} Due to the contribution of the signal polarization, under the proposed antenna element assignment only one antenna element could suffer in terms of zero RSP, and \underline{that antenna element is known}.
\end{enumerate}
As said, now we want to make the desired algorithms practically realizable. As a result, we first relax Assumption 1 by introducing background noise in the system with known average noise power $\sigma^2>0$. However, for an unknown environment, this power can be easily estimated \cite{n1, n2}. To relax Assumption 2, we bring in a decision threshold $K$ to check whether the received power at antenna $n$ is below or above this threshold. In  particular, after receiving $M$ samples of the received signal $x_{n}(m)$, the RSP at  the antenna element $n$ can be estimated as   
\begin{equation}
   P_n =\frac{1}{M} \sum_{m=1}^{M}|{x}_{n}(m)|^2.
   \label{RSPn}
\end{equation}
In the presence of a target, if $P_n < K$, we decide in the favor of the hypothesis

\begin{itemize}
    \item $H_0$: signal is not present at antenna element $n$ due to the compound steering element $a_n=0$ in (\ref{a_n}),
\end{itemize}
otherwise, we conclude hypothesis  
\begin{itemize}
    \item $H_1$: signal is present at antenna element $n$, and the compound steering element $a_n\neq 0$.
\end{itemize}

\begin{figure}[t!]
    \includegraphics[scale=0.59]{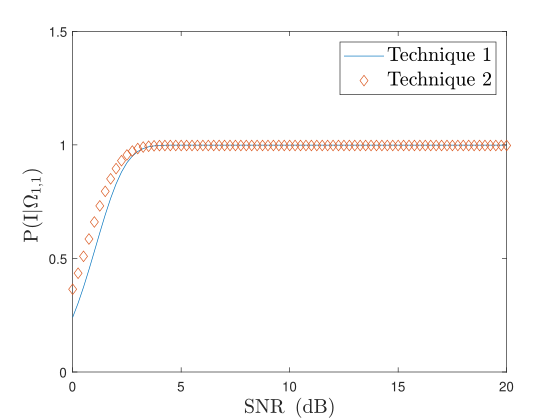}
    \caption{Probability of correctly identifying Event 1.}
    \label{fig:K1}
\end{figure}

Note that in low received signal-to-noise ratio (RSNR) (i.e., $\frac{|a_n|^2}{M}\sum_{m=1}^M \sigma^2_{\mathrm{s},m}/\sigma^2$) scenarios, RSP tests at multiple antennas may favor hypothesis $H_0$ for more than one antenna. If that is the case, our proposed antenna element alignment in Fig.~\ref{fig:Capture6} dictates us to discard the antenna output with the lowest measured RSP. Our analysis will shortly reveal that such a scenario occurs with negligible probability within the desired operating regime of RSNR.

Now, we discuss how to find the decision threshold $K$. A popular way of designing the threshold $K$ is based on the Neyman-Pearson lemma. Usually, a prefixed significance level $\alpha$ is set (e.g., $\alpha=0.001$) to restrict the probability of making a Type I error [i.e., $1-\mathrm{P}\paren{H_0| \Omega_{1,n}}$] to a certain percentage (in this case, 0.1\%). Next, a test is chosen to minimize Type II error [i.e., $1-\mathrm{P}\paren{H_1|a_n\neq 0}$].  Due to the monotonic nature of Type II error w.r.t. $K$, the value of $K$ will be determined by solving $ \mathrm{P}\paren{H_0| \Omega_{1,n}} = 1-\alpha$. Next, we present the derivation of  $ \mathrm{P}\paren{H_0| \Omega_{1,n}}$.

\subsection{Derivation of $ \mathrm{P}\paren{H_0| \Omega_{1,n}}$}
We propose two different techniques to derive $ \mathrm{P}\paren{H_0| \Omega_{1,n}}$ in order to find $K$. The first technique, namely Technique 1, is based on the probability density function (PDF) of $P_n$ and the second technique, namely Technique 2, applies the central limit theorem (CLT) to (\ref{RSPn}) and models $P_n$ as a Gaussian random variable. Now, we elaborate upon Technique 1.
\subsubsection{Technique 1}
Given $\Omega_{1,n}$, the received signal samples only contain noise. As a result, $|{x}_{n}(m)|^2$ in (\ref{RSPn}) is a sum of two squared independent identically distributed (i.i.d) Gaussian random variables each of which has a mean of 0 and a variance of $\sigma^2/2$. Thus, $|{x}_{n}(m)|^2$ is a central chi-square random variable with degrees of freedom $\rm{(DoF)} = 2$. As the received power $P_n$ in (\ref{RSPn}) is the average of $|{x}_{n}(m)|^2$ for $M$ samples, it also follows a central chi-square distribution with $\mathrm{DoF} = 2M$. The probability density function (PDF) of a central chi-square random variable with DoF = $\nu$ is \cite{chi} 
\begin{equation}
    f_{\nu}(w)=
\begin{cases}
0 & w\leq 0\,,\\
    \frac{w^{\frac{\nu}{2}-1}e^{-\frac{w}{2}}}{ 2^{\frac{\nu}{2}}\Gamma(\frac{\nu}{2})}&w > 0,\\
\end{cases}
    \label{chii}
\end{equation} 
\begin{figure}[t!]
    \includegraphics[scale=0.59]{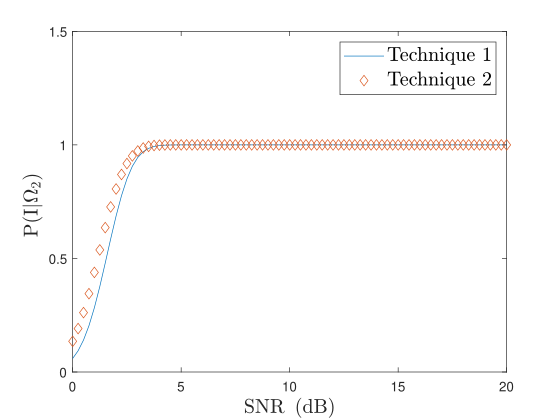}
    \caption{Probability of correctly identifying Event 2.}
    \label{fig:K2}
\end{figure}
and the corresponding cumulative distribution function (CDF) is
\begin{equation}
F_{\nu}(u)=\int_{0}^{u}f_{\nu}(w)\,dw =\begin{cases}
  0 & u\leq 0\,,\\
\frac{\Gamma_u (\frac{\nu}{2},\frac{u}{2})}{\Gamma (\frac{\nu}{2})} & u > 0\,,
\end{cases}
\label{CDF}
\end{equation}
where $\Gamma(\frac{\nu}{2})=\int_{0}^{\infty}t^{(\frac{\nu}{2}-1)}e^{-t}dt$ denotes the gamma function, and $\Gamma_u (\frac{\nu}{2},\frac{u}{2})=\int_{0}^{\frac{u}{2}}t^{(\frac{\nu}{2}-1)}e^{-t}\,dt$ represents the lower incomplete gamma function. Now, we use (\ref{chii}) to write the conditional PDF of the measured power $P_n$ given Event $\Omega_{1,n}$ as
\begin{equation}
    f_{P_n|\Omega_{1,n}}(w)=\frac{1}{\tilde{\sigma}^2}f_{\nu}\paren{\frac{w}{\tilde{\sigma}^2}}\,,
\label{PDFH0}
\end{equation} 
 where $\tilde{\sigma}^2=\sigma^2/2M$. Our goal was to find  $\mathrm{P}\paren{H_0| \Omega_{1,n}} =\mathrm{P}\paren{P_n\leq K | \Omega_{1,n}}$, which is
\begin{equation}
 \int_{0}^{K}  f_{P_n|\Omega_{1,n}}(w)\,dw=   \int_{0}^{K}\frac{1}{\tilde{\sigma}^2}f_{\nu}\paren{\frac{w}{\tilde{\sigma}^2}}dw =F_{\nu}\paren{\frac{K}{\tilde{\sigma}^2}}\,.
    \label{chi2}
\end{equation}
Last equality of (\ref{chi2}) is obtained from (\ref{CDF}). Now, we numerically solve $F_{\nu}\paren{\frac{K}{\tilde{\sigma}^2}}= 1-\alpha$ to find the decision threshold $K$. For fixed $M$ and $\sigma^2$, the value of $K$ can be precomputed. However, this method could be computationally complex and time consuming specially for a time-varying no\textit{}ise environment. That environment may demand a quick method and the following technique (i.e., Technique 2) could be found more useful.

\subsubsection{Technique 2}
To obtain a quick value of $K$, we apply CLT to (\ref{RSPn}). The chi-square distribution given in (\ref{PDFH0}) yields the conditional mean of $P_n$ as $\mu_{P_n|\Omega_{1,n}}= \sigma^2$, and its variance $\sigma^2_{P_n|\Omega_{1,n}}={\sigma^4}/M$.
The desired probability  $\mathrm{P}\paren{H_0| \Omega_{1,n}} = \mathrm{P}\paren{P_n\leq K | \Omega_{1,n}}$ which is
$$\mathrm{P}\paren{H_0| \Omega_{1,n}} = 1-\mathrm{P}\paren{P_n> K | \Omega_{1,n}},$$ and under CLT, it simplifies to
$$\mathrm{P}\paren{H_0| \Omega_{1,n}} =1- Q\paren{\frac{M\paren{K-\sigma^2}}{\sigma^4}}\,,$$ where $Q(\cdot)$ is the tail distribution function of the standard normal distribution. 
Now, we solve $Q\paren{\frac{M\paren{K-\sigma^2}}{\sigma^4}}= \alpha$ to get the value of $K$.
\begin{figure}[t!]
    \includegraphics[scale=0.59]{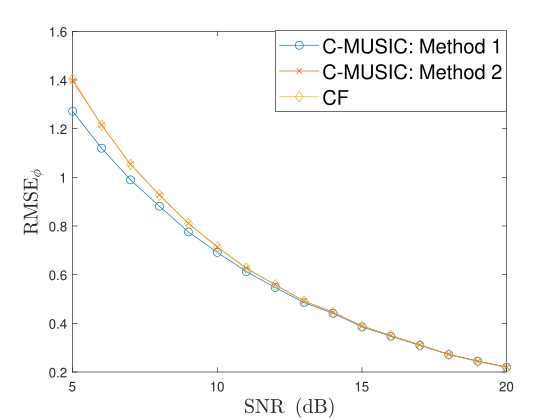}
    \caption{ $\rm{RMSE}_\phi$ vs average SNR (received) for Event 1.}
    \label{1p}
\end{figure}

\subsection{Analysis}
Recall that in a noiseless polarized sensitive environment, under the proposed antenna element assignment scheme two events can occur; Event 1 ($a_n=0$, where $n=1$ or 2 or 3 or 4), and  Event 2 ($a_n \neq 0$ for all $n$). Both C-MUSIC and CF operate differently in each of the above events. Therefore, the performance of the algorithms will depend on how accurately those events can be identified after processing $M$ samples of the received signal. Our next objective is to analyze $\mathrm{P}\paren{I|\Omega_i}$, which is the probability of  accurately identifying Event $i$ given Event $i$ occurred for a given set of source parameters, where $i=1,2$. Since Event 1 has 4 sub-events, and  $\mathrm{P}\paren{I|\Omega_1}$ is different for each of those sub-cases, we need to add another condition (i.e., $a_n=0$) in the derivation of the desired probability. Thus, for Event 1, we analyze $\mathrm{P}\paren{I|\Omega_{1,n}}$. In the derivation of the desired probability for Event 1, we will exploit the following two lemmas. The $\ist{1}$ one is straightforwardly obtained from (\ref{CDF}). 

\begin{lemma}
Given Event 1 due to $a_n=0$ (i.e., $\Omega_{1,n}$), the conditional probability of the measured received power at antenna $n$ (i.e., $P_{n}$) below the decision threshold $K$ is 
\begin{equation}
  \mathrm{P}\paren{P_n\leq K|\Omega_{1,n}}= F_{\nu}\paren{\frac{K}{\tilde{\sigma}^2}}\,.
\end{equation}
\label{lemmaho}
\end{lemma}
 Given event $\Omega_{1,n}$, the conditional PDF of the measured received power at antenna $n'\neq n$ follows a non-central chi-square distribution defined as \cite{ncchi,ncchi2}
\begin{equation}
   f_{P_{n'}|\Omega_{1,n}}(w)= \frac{1}{\tilde{\sigma}^2}f_{\nu,\,n'}\paren{\frac{w}{\tilde{\sigma}^2}}\,,
   \label{hiho}
\end{equation}
where
\begin{equation}
    f_{\nu,\,n'}(u)=
\begin{cases}
  \exp\paren{-\frac{\Delta_{n'}}{2}}\sum_{j=0}^{+\infty}\frac{\Delta^j_{n'}}{2^j j!}f_{2j+\nu}(u)&u > 0\,,\\
0&\text{otherwise.}
\end{cases}
    \label{chii0}
\end{equation}
\\
In (\ref{chii0}), $\Delta_{n'}=\sum_{m=1}^{M}2\mu^2_{m,\,n'}/\sigma^2$ is the non-centrality parameter where $\mu^2_{m,\,n'}=|{a}_{n'}|\sigma^2_{\mathrm{s},m}$  at the $\ith{n'}$ element, and $f_{2j+\nu}(u)$ follows the central chi-square distribution  (\ref{chii}) with $\mathrm{DoF} = 2j+\nu $. Integrating both sides of (\ref{chii0}) from $K$ to $\infty$, we get
\begin{equation}
     \bar{F}_{\nu,\,n'}(K) =  \exp\paren{-\frac{\Delta_{n'}}{2}}\sum_{j=0}^{+\infty}\frac{\Delta^j_{n'}}{2^j j!}\bar{F}_{2j+\nu}(K)\,,
\end{equation}
where $\bar{F}_{2j+\nu}(K) = 1-F_{2j+\nu}(K)$ is the complement of the CDF given in (\ref{CDF}). Now, we are ready to state the other lemma to get the desired probability $\mathrm{P}\paren{I|\Omega_{1,n}}$.
\begin{figure}[t!]
    \includegraphics[scale=0.59]{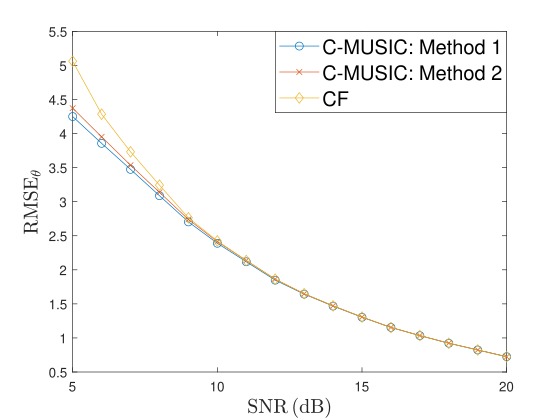}
    \caption{ $\rm{RMSE}_\theta$ vs average SNR (received) for Event 1.}
    \label{1t}
\end{figure}
\begin{lemma}
Given Event 1 due to $a_n=0$ (i.e., $\Omega_{1,n}$), the conditional probability of the received measured power at antenna $n'\neq n$ (i.e., $P_{n'}$) above the decision threshold $K$ is

\begin{equation}
    \mathrm{P}\paren{P_{n'}>K|\Omega_{1,n}} =  \bar{F}_{\nu,\,n'}\paren{\frac{K}{\tilde{\sigma}^2}}\,.
\end{equation}

\end{lemma}
Since the received noise vector at antenna $n$ is independent of that at antenna $n'\neq n$ for a given set of source parameters, we get the desired conditional probability $\mathrm{P}\paren{I|\Omega_{1,n}}$ as stated in the following theorem.
\begin{theorem}
Given Event 1 due to $a_n=0$ (i.e. $\Omega_{1,n}$), the conditional probability of  accurately identifying Event 1 is
\begin{eqnarray*}
  \mathrm{P}\paren{I|\Omega_{1,n}}&=& \mathrm{P}\paren{P_n\leq K|\Omega_{1,n}} \\
 && \times\prod_{n'=1,\, \neq n}^4 \mathrm{P}\paren{P_{n'}>K|\Omega_{1,n}}\\
  &=&  F_{\nu}\paren{\frac{K}{{\tilde{\sigma}^2}}}\prod_{n'=1,\, \neq n}^4 \bar{F}_{\nu,\,n'}\paren{\frac{K}{\tilde{\sigma}^2}} \,.
\end{eqnarray*}
   \label{chijjo}
\end{theorem}
\begin{figure}[t!]
    \includegraphics[scale=0.59]{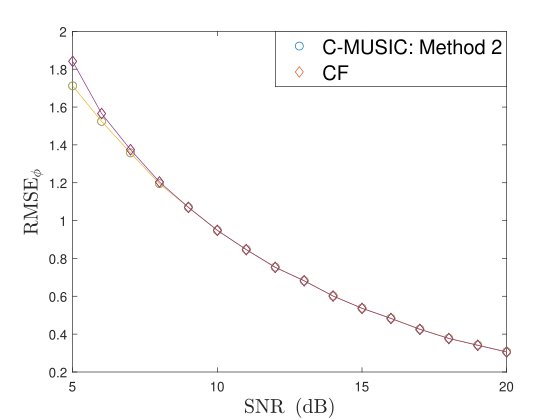}
    \caption{$\rm{RMSE_\phi}$ vs average SNR (received) for Event 2.}
    \label{2p}
\end{figure}
Reader must note that the probability in Theorem~\ref{chijjo} will change from one sub-event to another in Event 1. However, since $\bar{F}_{\nu,\,n'}\paren{\frac{K}{\tilde{\sigma}^2}} \leq 1$ and $ F_{\nu}\paren{\frac{K}{{\tilde{\sigma}^2}}} = 1-\alpha$, the probability in Theorem~\ref{chijjo} will have the same upper bound for all $n$ as given in the following corollary.
\begin{corollary}
Given Event 1 due to $a_n=0$ (i.e. $\Omega_{1,n}$), the conditional probability of  accurately identifying Event 1 is upper bounded by
\begin{eqnarray*}
  \mathrm{P}\paren{I|\Omega_{1,n}} \leq 1-\alpha; \quad \forall n\,.
\end{eqnarray*}
\label{coro}
\end{corollary}
In Event 2, $a_n\neq 0$ $\forall n$, and the conditional PDF of the measured received power $f_{P_{n}|\Omega_{2}}(w)$ follows a non-central chi-square distribution as shown in (\ref{hiho}). This distribution coupling with the fact that the received noise vector at antenna $n$ is independent of that at antenna $n'\neq n$ gives us the following theorem.

\begin{theorem}
Given Event 2 (i.e., $\Omega_{2}$), the conditional probability of  accurately identifying Event 2 is
\begin{eqnarray*}
  \mathrm{P}\paren{I|\Omega_{2}}&=& \prod_{n=1}^4 \mathrm{P}\paren{P_{n}>K|\Omega_{2}}= \bar{F}_{\nu,\,n}\paren{\frac{K}{\tilde{\sigma}^2}}\,.
 \end{eqnarray*}
    \label{chijjo2}
\end{theorem}
\section{Numerical Results}
In this section, our objectives are 1) to provide insights into the system performance through numerical examples based on our analytical results, and 2) to perform a thorough comparative study between C-MUSIC and CF algorithms. In all numerical examples, i) a 4-element UCA is considered with the antenna alignment scheme proposed in Figure~\ref{fig:Capture6}, ii) the number of samples $M=50$, iii) the significance level $\alpha=0.001$, iv) received waveform samples modelled as i.i.d complex Gaussian random variables with mean = 0 and variance = 1/2 per dimension, unless otherwise specified.  

\begin{figure}[t!]
    \includegraphics[scale=0.59]{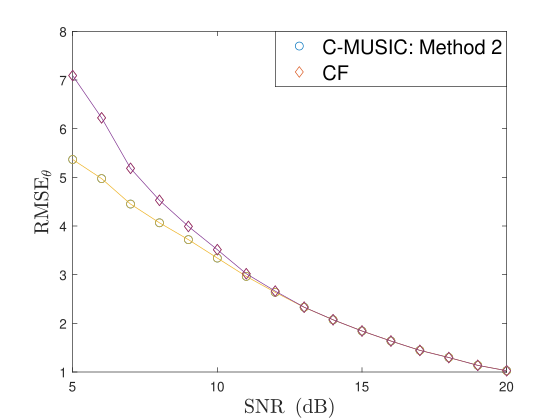}
   \caption{$\rm{RMSE_\theta}$ vs average SNR (received) for Event 2.}
    \label{2t}
\end{figure}

Recall that identification of each of the five effective events is performed employing the decision threshold $K$. It can be determined by setting Type I Error (i.e., $1- \mathrm{P}\paren{H_0|\Omega_{1,n}}$) equals to $\alpha$ using either Technique 1 or Technique 2. In Figure \ref{fig:K}, the decision threshold normalized by the average noise power is plotted by setting (Type I Error) $\alpha=0.001$. Here, it can be noticed that Technique 1 (which is based on the exact PDF of the decision statistics) yields a slightly higher value of $K$ than Technique 2 (which is based on the CLT). This higher value of $K$ will result in higher Type II error and its effects will be demonstrated shortly.

In the next example, we evaluate our derived expression of $\mathrm{P}\paren{I|\Omega_{1,n}}$ (i.e. the conditional probability of accurately identifying Event 1) given in Theorem~\ref{chijjo} as a function of average RSNR using both Technique 1 and Technique 2; see Figure \ref{fig:K1}. Here, Event 1 occurs due to the DOA  $\phi=45^{\circ}$, $\theta=10^{\circ}$, and polarization parameters $\eta=90^{\circ}$, $\gamma=45^{\circ}$. These parameters cause $a_n=0$ at $n=1$. In Figure \ref{fig:K1}, it can be noticed that Technique 2 slightly performs better than Technique 1 especially at low RSNR due to the use of a higher value of $K$. We also see that when the average RSNR is above or equal to 4 dB, both techniques exhibit the maximum achievable performance as given by the Corollary~\ref{coro}. In Figure \ref{fig:K2}, we plot $\mathrm{P}\paren{I|\Omega_{2}}$ (i.e. the conditional probability of  accurately identifying Event 2) given in Theorem~\ref{chijjo2} as a function of average RSNR for both Technique 1 and Technique 2. Here, the DOA and polarization parameters are $\phi=45^{\circ}$, $\theta=10^{\circ}$, $\eta=90^{\circ}$ and $\gamma=45^{\circ}$, respectively. Similar to Figure \ref{fig:K1}, the results in Figure \ref{fig:K2} suggest that both techniques are equally capable of identifying Event 2 almost perfectly when the average RSNR $\geq$ 4 dB. From now on, Technique 2 will be employed to design the decision threshold $K$.

Next, we would like to compare the performance of the CF algorithm against that of the C-MUSIC for Event 1 and Event 2. We first consider Event 1 due to $a_1=0$ (i.e., $\Omega_{1,1}$), where  $\phi=30^{\circ}$, $\theta=70.529^{\circ}$, $\eta=0^{\circ}$ and $\gamma=60^{\circ}$; see Figure \ref{1p} and \ref{1t}, where RMSE (Root Mean Square Error) is plotted as a function of the average RSNR. C-MUSIC has been implemented using both Method 1 and Method 2 where the former is observed to exhibit slightly better performance than the latter. Here, it can also be noticed that as expected, at low SNR (close to $5$ dB) the performance difference between C-MUSIC and CF algorithms is somewhat noticeable. However, as the average
RSNR increases that performance difference starts to diminish. At RSNR $\geq$ 10 dB, both proposed CF and C-MUSIC algorithms exhibit almost identical performance. Similar observations are made in Event 2 for $\phi=45^{\circ}$, $\theta=10^{\circ}$, $\eta=90^{\circ}$ and $\gamma=45^{\circ}$; see Figure \ref{2p} and \ref{2t}.

\begin{figure}[t!]
    \includegraphics[scale=0.59]{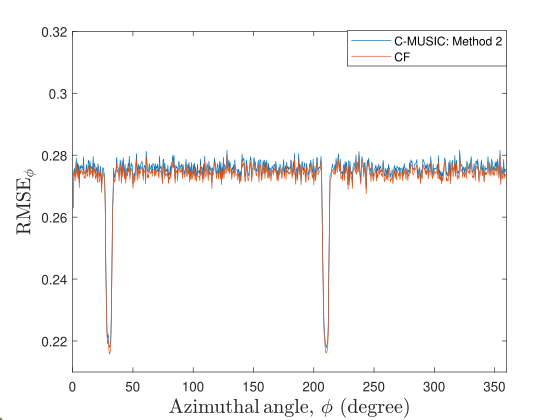}
    \caption{$\rm{RMSE_\phi}$ as a function of the azimuthal angle $\phi$.}
    \label{3p}
\end{figure}

\begin{figure}[t!]
    \includegraphics[scale=0.59]{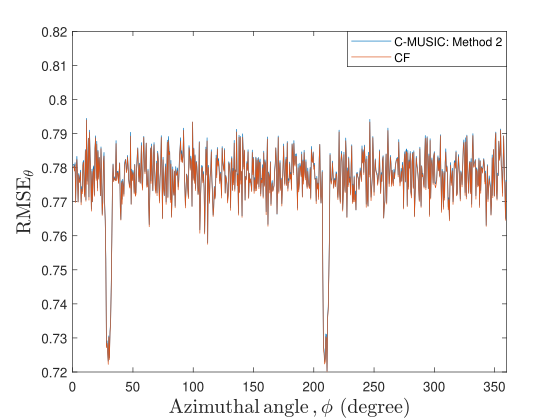}
   \caption{ $\rm{RMSE_\theta}$ as a function of the azimuthal angle $\phi$.}
    \label{3t}
\end{figure}

In the following numerical study, we would like to demonstrate the performance of both the algorithms as the system transits from one event to another due to the change in the azimuthal angle. Please see Figure \ref{3p}, and \ref{3t}, where we plot RMSE by varying the azimuth angle $\phi$ (with the increment of $\pm 0.5^{\circ}$). Here, we use SNR = $20$ dB, $\theta=70.529^{\circ}$, $\eta=0^{\circ}$ and $\gamma=60^{\circ}$. Note that when $\phi =30 ^{\circ}$, Event 1 occurs due to $a_1=0$ and otherwise, we have Event 2. Using the above two figures, we make the following set of observations: 1) the designed decision threshold $K$ is capable of efficiently differentiating between Event 1 and Event 2, 2) the RMSE of Event 2 is slightly higher than Event 1 since as $\phi$ approaches $30 ^{\circ}$, compound steering element $a_1$ approaches 0. Thus, it is better to discard the output of antenna element 1, 3) as expected, Event 1 occurs for two different azimuthal angles which are $180 ^{\circ}$ apart from each other, and 4) both the CF and C-MUSIC algorithms not only exhibit almost identical performance but also robust to the transition from Event 1 to Event 2 or vice versa. Similar results are also observed when we vary the elevation angles for a given azimuthal angle. For the conciseness, figures related to those observations are omitted.

Finally, we compare the complexity of the CF and C-MUSIC algorithms. In our calculation, we consider the complexity incurred after identifying Event $i$, where $i=1,2$. The complexities of the algorithms are measured by the number of real-time multiplications associated with the major operations. Note that one complex multiplication is equivalent to four real multiplications.
In C-MUSIC (Method 1), which is only applicable for Event 1, a new matrix is formed which requires $4(N-1)^2M+2(N-1)^2-(N-1)+2(N-1)^2-(N-1)$ real multiplications. Here, $4(N-1)^2M+2(N-1)^2-(N-1)$ is due to the estimation of the autocorrelation matrix and the additional $2(N-1)^2-(N-1)$ is an upper bound on the number of multiplications between the autocorrelation matrix and the diagonal matrix $\bd{F}_n$; $n=1,2,3,4$. As previously discussed, both C-MUSIC (Method 2) and CF operate by estimating the phases of the steering elements, $\kappa_1$, and $\kappa_2$. The cost associated with this estimation is $8M+2\log_2 p+1$ real multiplications for Event 1, and $16M+2\log_2 p+1$ real multiplications for Event 2, where $p$ refers to the number of digits of precision \cite{cost,bc}. It is known that the MUSIC algorithm performs eigenvalue decomposition (EVD) which often is obtained from a singular value decomposition (SVD). As per \cite{evd}, this complexity associated with an SVD is $12N^3$. As C-MUSIC (Method 1 and 2) operates on the 2D MUSIC algorithm, the DOA angle search using the null space requires $12N^3+N_{\theta}N_{\phi}\{4N(N-1)+2(N-1)+1\}$ real multiplications, where $N_{\theta}$ and $N_{\phi}$ represent the searching point number on the azimuthal and elevation planes, respectively. On the other hand, the required cost to estimate the DOA angles is upper bounded by $1+\log_2 p+3+1+\log_2 p$ from the estimates of $\kappa_1$, and $\kappa_2$ using the CF algorithm.  Here, the first two terms are related to the estimation of the azimuthal angle $\phi$, the third term is to calculate the $\kappa$, and the rest of the terms are associated with the estimation of the elevation angle $\theta$. All the corresponding costs are added and tabulated in Table \ref{t0} for complexity comparison. According to this table, the cubic order of the array size, the number of samples, and the product of the searching points dominate the complexity of the C-MUSIC algorithm, whereas, the complexity of the CF algorithm is primarily dictated by the number of samples.
\begin{table}[t]
\begin{center}
\caption{Complexity analysis}
\vspace{-2pt}
\begin{tabular}{|c|c|c|c|}
\hline
\multicolumn{2}{|c|}{Algorithm} & Event 1 & Event 2\\
\hline
\multirow{9}{*}{MUSIC} & \multirow{5}{*}{ Method 1} & $(N-1)^2(4M+4)$ &  \multirow{5}{*}{N/A}\\
 &  & $-2(N-1)$ &  \\
&  &$+12(N-1)^3$  &  \\
& & $+N_{\theta}N_{\phi}\times$ &\\
&  & $\{4N^2-10N+5\}$  &  \\
\cline{2-4}
 & \multirow{4}{*}{ Method 2} & $8M+2\log_2 p+1$  & $16M+2\log_2 p +1 $\\
&  & $+12(N-1)^3$ & $+12N^3 $\\
& &$+N_{\theta}N_{\phi}\times$ & $+N_{\theta}N_{\phi}\times$\\
&  & $\{4N^2-10N+5\}$ & $\{4N^2-2N-1\}$\\
\hline
\multicolumn{2}{|c|}{\multirow{2}{*}{CF}} & $8M+6$ & $16M+6$  \\
  \multicolumn{2}{|c|}{}   & $+4\log_2 p$ & $+4\log_2 p$\\
\hline
\end{tabular}
\label{t0}
\end{center}
\end{table} 

\begin{figure}[t!]
\centering
    \includegraphics[scale=0.59]{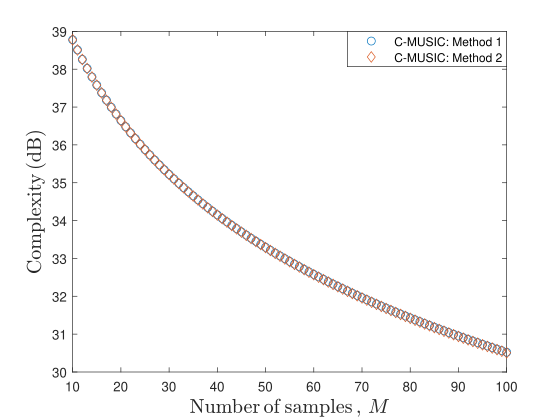}
    \caption{Complexity vs number of samples $M$ for Event 1.}
    \label{c2}
 \end{figure}
 \begin{figure}[t!]
    \includegraphics[scale=0.59]{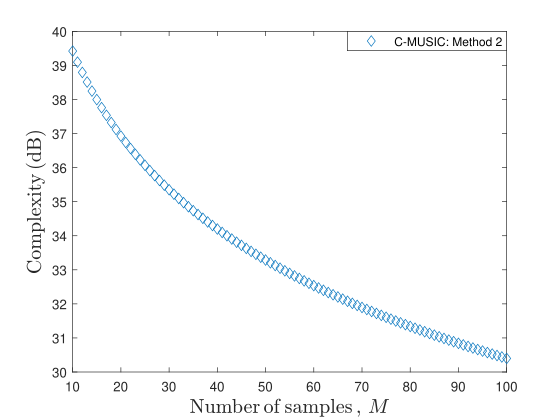}
   \caption{Complexity vs number of samples $M$ for Event 2.}
    \label{c1}
\end{figure}
Now, we use Table \ref{t0} to demonstrate the complexities of the DOA estimating algorithms by using numerical examples. Please see Fig. \ref{c2}, and \ref{c1}, where the complexities of C-MUSIC (w.r.t the CF algorithm) are plotted as functions of the number of samples $M$ for Event 1 and Event 2, respectively. Here, the search in MUSIC is conducted with $1^{\circ}$ precision of the DOA angles. The results in those figures jointly suggest that the CF algorithm offers significant complexity gain over the C-MUSIC. For instance, this gain is $33.29$ dB in Event 1 under Method 1, $33.28$ dB in Event 1 under Method 2 and $33.29$ dB in Event 2 for $M=50$. This gain increases as the number of samples $M$ decreases.

\section{Conclusions}
In this paper, we addressed the problem of localizing a single narrowband source in all possible polarization scenarios just employing simple (short dipole) antenna elements and signal processing techniques. Depending on the antenna alignment, the contribution of the polarization and DOA angles could result in poor received signal power at one or more antenna elements. To overcome this issue, an antenna alignment scheme was mathematically developed for a UCA that operates with the minimum number of required antenna elements. Under this scheme, antenna elements are aligned in such a way that not more than one element will suffer from low received signal power due to the joint effects of polarization and DOA angles. A decision threshold was developed to decide whether the antenna element with the smallest received power should be considered in the process of DOA estimation or not. We demonstrated how the polarization contribution can be suppressed from the non-signal subspace in order for the popular MUSIC algorithm to operate in all the polarization scenarios. During the process of cleaning the non-signal subspace, we designed an algorithm that estimates the DOA angles in a closed-form manner. Of course, the latter is significantly less complex than the former. Our numerical results demonstrated that depending on the system condition, that gain could be up to 32.29 dB without sacrificing any performance as long as the received SNR is 10 dB or more. 
 \bibliographystyle{ieeetran}
\bibliography{Reference.bib}
\vspace{12pt}

\end{document}